\documentclass[review,12pt]{elsarticle}

\usepackage{amssymb}
\usepackage{amsthm}
\usepackage{amsmath}

\usepackage{mathrsfs}
\usepackage{graphicx}
\usepackage{epstopdf}
\usepackage{float}
\usepackage{caption}
\usepackage{subcaption}
\usepackage{bm}
\usepackage{bbm}
\usepackage{mathrsfs}
\usepackage{cleveref}
\usepackage{soul}
\usepackage{accents}
\usepackage{graphicx}
\usepackage{xcolor}
\usepackage{courier} 
\usepackage{psfrag} 
\usepackage{listings} 
\usepackage{tabu} 
\usepackage{longtable}
\usepackage{changepage} 
\usepackage[margin=2.2cm]{geometry}
\biboptions{sort&compress} 

\journal{Continuum Mechanics and Thermodynamics}

\makeatletter
\def\@author#1{\g@addto@macro\elsauthors{\normalsize%
    \def\baselinestretch{1}%
    \upshape\authorsep#1\unskip\textsuperscript{%
      \ifx\@fnmark\@empty\else\unskip\sep\@fnmark\let\sep=,\fi
      \ifx\@corref\@empty\else\unskip\sep\@corref\let\sep=,\fi
      }%
    \def\authorsep{\unskip,\space}%
    \global\let\@fnmark\@empty
    \global\let\@corref\@empty  
    \global\let\sep\@empty}%
    \@eadauthor={#1}
}
\makeatother

\begin{document}

\begin{frontmatter}



\title{Analysis of hydrogen diffusion in the three stage electro-permeation test}


\author[Cam]{Arun Raina}

\author[Cam]{Vikram S. Deshpande}

\author[IC]{Emilio Mart\'{\i}nez-Pa\~neda}

\author[Cam]{Norman A. Fleck\corref{cor1}}
\ead{naf1@cam.ac.uk}

\address[Cam]{Department of Engineering, Cambridge University, CB2 1PZ, Cambridge, UK}

\address[IC]{Department of Civil and Environmental Engineering, Imperial College London, London SW7 2AZ, UK}

\cortext[cor1]{Corresponding author.}

\begin{abstract}
The presence of hydrogen traps within a metallic alloy influences the rate of  hydrogen diffusion. The electro-permeation (EP)  test can be used to assess this: the permeation of hydrogen through a thin metallic sheet is measured by suitable control of hydrogen concentration on the front face and by recording the flux of hydrogen that exits the rear face. Additional insight is achieved by the more sophisticated three stage EP test: the concentration of free lattice hydrogen on the front face is set to  an initial level, is then dropped to a lower intermediate value and is then restored to the initial level. The flux of hydrogen exiting the rear face is measured in all three stages of the test. In the present study, a transient analysis is performed of hydrogen permeation in a three stage EP test, assuming that lattice diffusion is accompanied by trapping and de-trapping. The sensitivity of the three stage EP response to the depth and density of hydrogen traps is quantified. A significant difference in permeation response can exist between the first and third stages of the EP test when the alloy contains a high number density of deep traps.
\\
\end{abstract}


\end{frontmatter}


\section{Introduction}
\label{intro}

There is growing interest in the use of high strength alloys for structures and components exposed to hydrogen-containing environments, such as sub-sea structures for oil transport or pressure vessels for hydrogen storage \cite{Gangloff2012,Djukic2019}. However, dissolved hydrogen can degrade the ductility and fracture toughness of high strength alloys \cite{AM2016,Shishvan2020}. The susceptibility of metallic structures to hydrogen-assisted failure is sensitive to the solubility and rate of diffusion of hydrogen atoms in the crystal lattice \cite{Ayas2014,JMPS2020}. Microstructural defects such as dislocations, grain boundaries and vacancies act as hydrogen `traps' that sequester hydrogen and retard diffusion \cite{pundt+kirchheim2006,bhadeshia1}. 

A physically motivated model for the diffusion of hydrogen in metallic alloys has been formulated by McNabb and Foster \cite{mfoster63} for the diffusion of hydrogen in an alloy where the density of traps is lower than the density of lattice hydrogen sites, see for example \cite{raina+etal17a}. McNabb and Foster \cite{mfoster63} account for the kinetics of trapping and detrapping from a single type of microstructural defect (trap) uniformly distributed in the alloy. Coupled partial differential equations are derived for the concentrations of lattice and trapped hydrogen in space and time \cite{mfoster63}, and numerical solutions have been obtained for these equations   \cite{Legrand2015,CS2020b}. The McNabb and Foster model requires rate constants for the trap kinetic equations, and the experimental determination of these is complicated. Oriani \cite{oriani70} simplified the analysis by assuming that hydrogen atoms at lattice and trap sites are in local equilibrium due to very fast trapping and detrapping. The local equilibrium condition implies that the differential equation for trap kinetics is replaced by an algebraic relation between the hydrogen concentration in the lattice and in the traps.


Thermal desorption spectrometry (TDS) and electrochemical permeation (EP) experiments are commonly used to measure the characteristics of hydrogen diffusion and trapping in alloys, including trap densities and trap binding energies (and thereby the type of traps) \cite{Li2004,Frappart2010,Zafra2022}. In a TDS test, the temperature of a hydrogen-containing sample is increased at a fixed rate and the rate of hydrogen desorption from the sample is measured. In contrast, EP tests are performed at a fixed temperature and a fixed hydrogen concentration is imposed on the front, hydrogen-inlet side of a plate-like specimen whereas a vanishing hydrogen concentration is imposed on the rear, hydrogen-outlet side of the specimen, of thickness $L$.  In a single stage test, the hydrogen concentration is initially very low in the specimen, and upon imposing a finite fixed concentration on the inlet side, the rate of hydrogen efflux from the outlet side increases in a transient manner until it achieves a steady state value.   Recently, Raina and co-workers conducted an analytical and numerical analysis of TDS \cite{raina+etal17b} and of single-stage EP \cite{raina+etal17a,raina+sime2018} experiments. The regimes of behaviour of hydrogen diffusion were mapped out, producing design maps that serve as guidelines for a determination of trap characteristics.

Three stage EP tests \cite{turnbull+etal89,ha+ai+scully14,Kim2014,fallahmohammadi+etal2013} give additional insight into lattice diffusivity in the presence of deep traps, and reveal the effect of irreversible traps upon hydrogen transport.  \emph{Irreversible} or \emph{deep} traps are those associated with a large enthalpy of trapping in relation to the activation energy for diffusion of hydrogen in the lattice. On the other hand, \emph{reversible} or \emph{shallow} traps have enthalpies comparable to the self-activation energy for hydrogen diffusion in the lattice.

\begin{figure}[htp]
\centering
\includegraphics[width=0.8\textwidth]{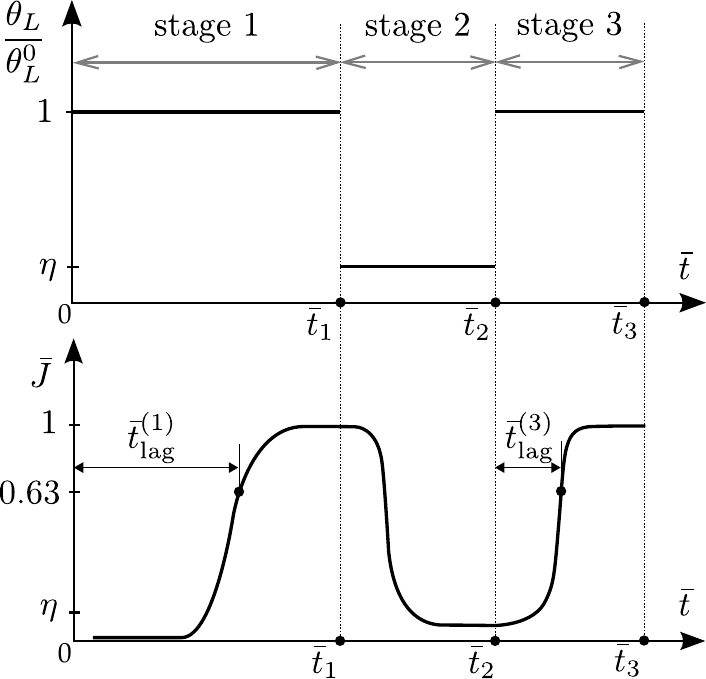}
\caption{The three stage EP test with the boundary condition at \( \bar{x}=0 \) throughout the test and the corresponding hydrogen flux. Stage 1 attains a steady state at time \( {{\bar{t}}_{1}} \); stage 2 starts at \( {{\bar{t}}_{1}} \) and continues until steady state at \( {{\bar{t}}_{2}} \); and finally stage 3 starts at \( {{\bar{t}}_{2}} \) and achieves a steady state at \( {{\bar{t}}_{3}} \). The time lags  to achieve steady state in stages 1 and 3 are denoted by \( \bar{t}_{\text{lag}}^{(1)} \) and \( \bar{t}_{\text{lag}}^{(3)} \), respectively; both time lags are identified by the instant when the normalised flux  $\bar{J}$ attains a value of 0.63. The parameter $\eta$  (such that \( 0\le \eta \le 1 \)) is the ratio of lattice occupancy applied in stage 2 to that applied in stage 1, at $\bar{x}=0$. Steady state in stage 2 is identified by the flux \( \bar{J}\to \eta \).}
\label{fig:epstages}
\end{figure}

The three stage EP test \cite{turnbull+etal89,ha+ai+scully14,Kim2014,fallahmohammadi+etal2013} is described in Fig. \ref{fig:epstages}. The test is performed in accordance with the ASTM standard G148-97(2011) using the double-cell apparatus proposed by Devanathan and Stachurski \cite{devanathan1964}.  In the first stage, a fixed value of charging current density enforces a constant concentration of hydrogen on the front side of the plate-like specimen. Hydrogen permeates through the specimen and a rising flux of hydrogen is measured at  the exit side until a steady state value is attained. In the second stage, the charging current is switched to a very low value (background current) in order to ensure that a low value of free, lattice hydrogen is attained throughout the thickness of the specimen. Finally, in the third stage, the same current density as that in the first stage is applied, and the transient flux of hydrogen is measured on the exit side until a steady state is again established. It is generally assumed that the irreversible traps are full at the end of the first stage such that only lattice hydrogen and reversible traps are operative in the second and third stages. The aim of the present study is to explore the sensitivity of hydrogen diffusion in a three stage EP test to the presence of traps.  

The remainder of this study is organised as follows. In Section \ref{sec2}, the theory of hydrogen diffusion in a metallic alloy is summarised. The governing partial differential equation (PDE) for one dimensional lattice diffusion in the presence of traps is presented in terms of suitable non-dimensional groups. The boundary conditions and initial state for solving the governing PDE in the three stage EP test are given. In Section \ref{sec3}, the main results from an asymptotic analysis of the first stage of an EP test are summarised from \cite{raina+etal17a}. Finally, in Section \ref{sec4}, numerical simulations of the three stage EP tests are presented.

\section{Theory of diffusion of hydrogen in a 3 stage EP test}\label{sec2}

\subsection{Kinetics of hydrogen diffusion}\label{sec2a}
Hydrogen atoms occupy normal interstitial lattice sites (NILS) and, additionally, can  reside at trapping sites such as interfaces or dislocations. The total hydrogen concentration \( C \)  is the sum of lattice hydrogen concentration \( {{C}_{L}} \)  and trapped hydrogen concentration $C_T$. The distribution of lattice concentration \( {{C}_{L}}(x,t) \) in space \( x \) and time \( t \) is dictated by Fickian diffusion over the NILS. However, when the lattice contains traps, hydrogen diffusion is modified by both trapping and detrapping of hydrogen atoms \cite{mfoster63}. Mass conservation dictates that the rate of change of total concentration equals the net flux of diffusing hydrogen atoms \( {{C}_{L}}(x,t) \) and, in one-dimensional form, this can be written as
\begin{equation}
\frac{\partial \,{{C}_{L}}}{\partial t}+\frac{\partial \,{{C}_{T}}}{\partial t}={{D}_{L}}\frac{{{\partial }^{2}}{{C}_{L}}}{\partial {{x}^{2}}}.
\label{eq1}
\end{equation}
Here, \( {{D}_{L}}={{D}_{0}}\text{exp}(-Q/RT) \)  is the lattice diffusion coefficient and is expressed in terms of the temperature \( T \), lattice activation energy \( Q \), diffusion pre-exponential factor \( {{D}_{0}} \) and the universal gas constant \( R \).

It is convenient to introduce the lattice and trap occupancy fractions \( \theta_{L} \)  and \( {{\theta}_{T}} \) , respectively, by re-writing the lattice and trap concentrations in the form \( {{C}_{L}}={{\theta }_{L}}\,\beta \,{{N}_{L}} \) and $C_T=\theta_T\,\alpha\,N_T$. Here, \( \beta \)  is the number of NILS per lattice atom, \( {{N}_{L}} \) is the number of lattice atoms per unit volume, \( \alpha \)  is the number of atom sites per trap and \( {{N}_{T}} \)  is the number of trap sites per unit volume. We emphasise that \( 0\le {{\theta }_{L}}\le 1 \)  and \( 0\le {{\theta }_{T}}\le 1 \). Using these relations and assuming that the number of trap sites remains constant, Eq. (\ref{eq1}) can be re-written as 
\begin{equation}
\frac{\partial \,{{\theta }_{L}}}{\partial t}+\left( \frac{\alpha {{N}_{T}}}{\beta {{N}_{L}}} \right)\frac{\partial \,{{\theta }_{T}}}{\partial t}
={{D}_{L}}\frac{{{\partial }^{2}}{{\theta }_{L}}}{\partial {{x}^{2}}}.
\label{eq2}
\end{equation}
The net rate of trapped hydrogen concentration \( \partial {{\theta }_{T}}/\partial t \)  is obtained by considering the kinetics of trapping and detrapping, using standard rate theory \cite{devanathan1964}, summarised as follows. 

Oriani \cite{oriani70} assumed that the rate of trapping and detrapping is sufficiently fast for local equilibrium to exist between the hydrogen atoms at lattice sites and at trap sites \cite{raina+etal17b}, such that,
\begin{equation}
    \frac{\theta_T}{1-\theta_T} = K \frac{\theta_L}{1-\theta_L} 
\end{equation}
For the practical case where \( {{\theta }_{L}}\ll 1 \), we have 
\begin{equation}
\theta_T = \dfrac{K\theta_L}{1+ K\theta_L}
\label{eq3}
\end{equation}
where the equilibrium constant \( K \)  is given in terms of the trap binding energy \( \Delta H \)  as 
\begin{equation}
K=\text{exp}\left\{ \frac{-\Delta H}{RT} \right\}.
\label{eq4}
\end{equation}
The Oriani assumption of local equilibrium has been explored by Raina et al. \cite{raina+etal17b} (see their Supplementary Information). They show that local equilibrium is maintained for both trapping and detrapping when the hopping rate is of realistic magnitude. Recall that the temperature \( T \)  is held constant in EP tests.

\subsection{Governing diffusion equation with local equilibrium}\label{sec2b}

Assume that local equilibrium exists between hydrogen atoms at the lattice and the trap sites  \cite{raina+etal17a}. Then, upon making use of \eqref{eq3}-\eqref{eq4}, the PDE \eqref{eq2} can be re-cast in the simpler form 
\begin{equation}
 \frac{\partial \,{{\theta }_{L}}}{\partial t}\left[ 1+\frac{\alpha {{N}_{T}}K}{\beta {{N}_{L}}{{(1+K{{\theta }_{L}})}^{2}}} \right]
={{D}_{L}}\frac{{{\partial }^{2}}{{\theta }_{L}}}{\partial {{x}^{2}}} ,
\label{eq6}
\end{equation}
%
%
%
Identification of an appropriate set of non-dimensional groups leads to a much simpler statement of the governing PDE as follows. 

\subsection{Non-dimensional groups and governing PDE}\label{sec2c}

Consider a one-dimensional specimen of length \( L \) spanning the domain \( 0\le x\le L \). Introduce the following non-dimensional quantities: spatial coordinate \( \bar{x}\equiv x/L \), time \( \bar{t}\equiv t{{D}_{L}}/{{L}^{2}} \), trap density \( \bar{N}\equiv (\alpha {{N}_{T}})/(\beta {{N}_{L}}) \), lattice activation energy \( \bar{Q}\equiv Q/(R{T}) \), trap binding energy \( \Delta\overline{ H}\equiv \Delta H/(R{T}) \) and fractional lattice occupancy \( {{\bar{\theta }}_{L}}={{\theta }_{L}}/\theta _{L}^{0} \), where  \( \theta_{L}^{0} \)  is the initial lattice occupancy fraction.  Then, equation \eqref{eq6} reduces to 
\begin{equation}
\dfrac{\partial{\bar\theta}_L}{\partial\bar{t}}
\bigg[1+\dfrac{K\bar{N}}{\big(1+((K\theta^0_L)^2(\bar\theta_L)^2)}\bigg]  =\dfrac{\partial^2\bar\theta_L}{\partial\bar{x}^2} \,.
\label{eq9}
\end{equation}
It is evident that the PDE delivers \( {{\bar{\theta }}_{L}} \) as a function of \( \bar{x}\) at any time \( \bar{t}\), for any assumed values of $K\bar{N}$ and $K\theta^0_L$. This PDE can be simplified and solved for extreme choices of $K\bar{N}$ and $K\theta^0_L$, as discussed below.

\subsection{Initial condition and boundary conditions for the three stage EP test}\label{sec2d}
Numerical simulations are performed of the three stage EP test, with two rise transients in stages 1 and 3 \cite{turnbull+etal89}. Consider the one-dimensional EP experiment with hydrogen introduced into an initially hydrogen-free specimen of length \( L \) at time \( t=0 \) at the front face of the specimen, \( x=0 \). The lattice occupancy fraction at the exit face \( x=L \) during all stages of the test is maintained at zero.

The scope of the present study is limited to the case where the rate of permeation is controlled by hydrogen diffusion through the alloy rather than the rate of surface reactions that produce the hydrogen and the rate of hydrogen transfer into the alloy. The test procedure in the ASTM G148-97(2011) standard demands that the thickness of the alloy is sufficient to ensure that diffusion controls the rate of permeation. It is also assumed that the rate of surface reaction is taken to be sufficiently high that a concentration boundary condition is appropriate rather than a flux boundary condition.
Although the term galvanostatic permeation test is commonly used for electro-permeation tests, the flux of hydrogen through the alloy specimen is often limited by diffusion with a boundary condition of concentration on the entry side rather than a prescribed flux of hydrogen into the specimen. It is straightforward to perform a numerical analysis of hydrogen permeation with other boundary conditions enforced, such as a prescribed entry flux of hydrogen see for example Pumphrey \cite{Pumphrey1980}. But this is beyond the scope of the present study. 

In the first stage of the three stage EP test, the occupancy of lattice hydrogen is maintained at \( \theta _{L}^{0} \) at \( \bar{x}=0 \) until a steady state efflux is achieved at time \( {\bar{t}_{1}} \). The second stage starts at time \( {\bar{t}_{1}} \), and the charging condition at \( \bar{x}=0 \) is dropped to \( \eta \theta_{L}^{0} \). The fraction $\eta$ dictates the amount of lattice hydrogen that diffuses out of the specimen in the second stage. We restrict the range of $\eta$ to $\eta\in[0.001,\,0.1]$. The second stage continues until the flux has dropped to a  steady state value at time \( {\bar{t}_{2}} \). It is commonly assumed that, at the end of the second stage,  the residual concentration of lattice hydrogen is negligible while deep traps remain full. The third stage of the test starts at time \( {\bar{t}_{2}} \) and the charging condition at \( \bar{x}=0 \) is again set to \( \theta_{L}^{0} \); the efflux increases  until a steady state is attained at time \( {\bar{t}_{3}} \). 
 

The boundary conditions and initial conditions to solve  \eqref{eq9} are:
\begin{itemize}
\item {\bf Stage 1:}
\begin{equation}
 \bar\theta^{(1)}_{L} = 
  \begin{cases} 
   1 \,\, {:} &  \bar{x} = 0 \\
   0 \,\, {:} &  \bar{x} = 1
 \end{cases}\quad{\rm and}\quad 
 \theta^{(1)}_{L}(\bar{x},\bar{t}=0)=0
 \label{eq10}
\end{equation}

\item {\bf Stage 2:}
\begin{equation}
 \bar\theta^{(2)}_{L} = 
  \begin{cases} 
   \eta  \,\, {:} &   \bar{x} = 0 \\
   0    \,\, {:} & \bar{x} = 1
 \end{cases}\quad{\rm and}\quad
 \theta^{(2)}_{L}(\bar{x},\bar{t}=\bar{t}_1)=\theta^{(1)}_{L}(\bar{x},\bar{t}_1)
 \label{eq11}
\end{equation}

\item {\bf Stage 3:}
\begin{equation}
 \bar\theta^{(3)}_{L} = 
  \begin{cases} 
   1     \,\, {:} & \bar{x} = 0 \\
   0     \,\, {:} & \bar{x} = 1
 \end{cases}\quad{\rm and}\quad
 \theta^{(3)}_{L}(\bar{x},\bar{t}=\bar{t}_2)=\theta^{(2)}_{L}(\bar{x},\bar{t}_2)
 \label{eq12}
\end{equation}
\end{itemize}
where the fractional occupancy fraction \( {{\bar{\theta }}_{L}^{(i)}} \)  corresponds to each stage \textit{i}. 

The flux of hydrogen \( J \)  measured at the exit face (\( x=L \) ) is given by
\begin{equation}
J(t) = \bigg(\dfrac{-D_L\beta N_L \theta^0_L}{L}\bigg)\dfrac{\partial\bar\theta_L}{\partial\bar{x}}
\label{eq13}
\end{equation}
This flux \( J \) represents the number of hydrogen atoms exiting the specimen per unit area per unit time. In the following, we shall present a normalised flux \( \bar{J}=J/{{J}_{ss}} \) where \( {{J}_{ss}} \) is the steady-state flux given by
\begin{equation}
{{J}_{ss}}=\frac{{{D}_{L}}\,\beta {{N}_{L}\theta_{L}^{0}}}{L}.
\label{eq14}
\end{equation}
We define the time lags $\bar{t}^{(1)}_{\text{lag}}$ and $\bar{t}^{(3)}_{\text{lag}}$ in stages 1 and 3, respectively, as the time required for the flux to attain the value \( \bar{J}=0.632 \) from the start of each stage \cite{barrer41}. Recall that a sketch of the boundary conditions in the  three stage EP test, with resulting efflux transients is given in Fig. \ref{fig:epstages}.

\section{Overview of the asymptotic analysis in Stage 1}\label{sec3}
\begin{figure}[htp]
\centering
\includegraphics[width=0.9\textwidth]{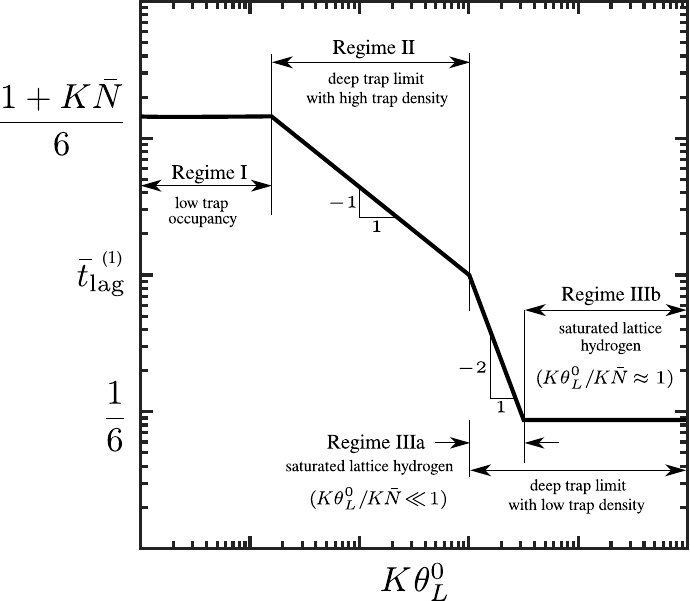}
\caption{Sketch of the time lag $\bar{t}^{(1)}_{\text{lag}}$ versus \(  K \theta_{L}^{0} \), to illustrate the four regimes of behaviour as deduced from the asymptotic analysis of the EP test in \cite{raina+etal17a}. The regimes are marked on the map for a fixed value of $K \bar{N}$. Modified from  \cite{raina+etal17a}.}
\label{fig4}
\end{figure}

In order to help guide the interpretation of numerical results, we first recall the regimes of behaviour by asymptotic analysis of the single-stage EP test \cite{raina+etal17a}. These relations, formulated in terms of \( {{\bar{t}}_{\text{lag}}} \), also inform the permeation transients in stage 3 of EP tests. The time lag values given in this section corresponds to the time lag \( \bar{t}_{\text{lag}}^{(1)} \) in stage one but the superscript 1 is dropped for conciseness. As shown in Fig. \ref{fig4}, four distinct regimes of behaviour are identified. \\ 

\noindent\textbf{Regime I:} The shallow trap  limit with \( K\theta_{L}^{0}\ll 1 \). In this regime, traps have low occupancy and diffusion takes place with an effective diffusion coefficient \( {D}_{\text{eff}}={{D}}_{L}/(1+K\bar{N}) \). The time lag \( {{\bar{t}}_{\text{lag}}} \) for hydrogen to reach \( \bar{x}=1 \) is given by \cite{raina+etal17a}
\begin{equation} 
{}{{\bar{t}}_{\text{lag}}}=\frac{1+K\bar{N}}{6}.
\label{eq18}
\end{equation} 

\noindent\textbf{Regime II:} The deep trap limit with \( K\theta _{L}^{0}\gg 1 \) and high trap density such that \( \dfrac{{K\bar{N}}}{(K{{\theta _{L}^{0})}^{2}}}\gg 1 \). In this case, the traps are full.  The solution as derived in \cite{raina+etal17a}  reads
\begin{equation}
{{\bar{t}}_{\text{lag}}}=\frac{1}{2}\bigg( \frac{{K\bar{N}}}{K\theta_L^0} \bigg).
\label{eq22}
\end{equation}

\noindent\textbf{Regime III:} The deep trap limit with $K\theta _{L}^{0}\gg 1$ but low trap density such that \( \dfrac{{K\bar{N}}}{(K{{\theta _{L}^{0})}^{2}}}\le 1 \). The traps are again full with \( {{\theta }_{T}}=1 \).
Two sub-regimes are identified in \cite{raina+etal17a}:

\textit{Case a} for which $K\theta^0_L/K\bar{N}\ll 1$. Here, 
\begin{equation}
\bar{t}_{\text{lag}} =\dfrac{\pi}{4}\bigg(\dfrac{K\bar{N}}{K\theta^0_L}\bigg)^2
\label{eq25}
\end{equation}

\textit{Case b} for which $K\theta^0_L/K\bar{N}\ \approx 1$. Here, 
\begin{equation}
{{\bar{t}}_{\text{lag}}}=\frac{1}{6}.
\label{eq26}
\end{equation}
Note that, apart from the diffusion in regime IIIb, the lag time for diffusion of hydrogen  is affected by the trap characteristics \( (K\bar{N}, K \theta _{L}^{0})\). 


The regime map shown in Fig. \ref{fig4} is based on equations \eqref{eq18}, \eqref{eq22}, \eqref{eq25} and \eqref{eq26} corresponding to regimes I, II, IIIa and IIIb, respectively. For a given \( K \bar{N} \), the response transitions from regime I to II to IIIa and finally regime IIIb with increasing \( K \theta _{L}^{0} \).  At a very low value of \( K \theta_{L}^{0} \) the trap occupancy is negligible and the response lies in regime I with \( {{\bar{t}}_{\text{lag}}} \) independent of \( K \theta_{L}^{0} \). With increasing \( K \theta_{L}^{0} \) the behaviour transitions to regime II and in this second regime \( {{\bar{t}}_{\text{lag}}} \) is inversely proportional to \( K \theta_{L}^{0} \). A further increase in \( K \theta_{L}^{0} \) results in regime IIIa and \( {{\bar{t}}_{\text{lag}}} \) scales inversely with \( {{\left( K \theta_{L}^{0} \right)}^{2}} \). Finally, at large values of \( K \theta_{L}^{0} \), regime IIIb is entered and \( {{\bar{t}}_{\text{lag}}} \)  again becomes independent of \( K \theta_{L}^{0} \); however, in contrast to regime I,  Fickian diffusion occurs within the lattice.\\

The diffusion model of the current study can be extended in a straightforward manner when the alloy contains multiple traps of varying density and binding energy. Assume that diffusion occurs within the lattice and that each type of trap acts independently from its neighbours. The Oriani relation (3) is used to determine the density of each type of trap in terms of its binding energy and lattice occupancy. Additionally, the above analytical formulae for the time lag in a single stage EP test and single type of trap can be modified in a straightforward manner for multiple traps provided that each type of trap has a density and binding energy that places the response in the same regime I to III. For example, consider an alloy that contains $n$ types of trap, such that for each type $i$ of trap has an equilibrium constant $K_i$, number of traps $N_i$ per unit volume, and number of hydrogen atom sites $\alpha_i$. The non-dimensional trap density for this trap is $\bar{N}_i = \alpha_i N_i / (\beta N_T )$.  Then if all of the traps are shallow such that $K_i \theta_L^0 << 1$ the alloy has a lag time for hydrogen to reach $\bar{x}=1$ of
\begin{equation}
    \bar{t}_{\text{lag}} = \frac{1+\sum_{i=1}^n K_i N_i}{6}
\end{equation}
and regime I behaviour is maintained when the single type of trap is replaced by multiple traps. In similar manner, the expressions for the time lag in regimes II and III can be generalised for multiple traps provided each trap, when existing alone, gives a response in the same regime.\\ 
There is no simple extension of the analytical formulae for the lag time when the alloy contains traps of varying density and binding energy such that each type of trap when present alone gives a different regime of response. Although full numerical predictions can still be made, the number of parameters that are introduced by the presence of multiple traps is sufficiently large that it is prohibitive to give a complete overview of the response over all of parameter space. Instead, a restricted set of numerical simulations can be performed for an assumed distribution of traps. This approach has been explored for single stage EP tests by Kirchheim \cite{Kirchheim2016}, for example.  The present study is a first attempt to identify the effect of trap density and binding energy upon the relative time lag in the first and third stages of a 3 stage EP test. Accordingly, we consider the idealised case of a single population of traps of fixed binding energy rather than a distribution of traps. It is recognised that a distribution of trap binding energies can exist for commercial alloys but it is beyond the scope of the present preliminary study to address this more complex case involving a specific choice of a density distribution of multiple traps.

\section{Numerical analysis of three stage EP tests}\label{sec4}
Theoretical insight is now gained into hydrogen transport in the three stage EP test  by solving the governing PDE \eqref{eq9}, subject to the appropriate initial and boundary conditions (see Section \ref{sec2d}). The PDE \eqref{eq9} is solved numerically by using the partial differential equation solver \textit{pdepe} in MATLAB\footnote{The pdepe solver is based on the method of lines which converts the given PDE into a system of initial value problems. In this method, the spatial derivatives are replaced with algebraic approximations and the remaining time derivatives are solved as a system of ordinary differential equations. An automatic time-stepping routine in pdepe solver ensures temporal convergence is achieved in each solution step. All simulations used a uniform mesh with element size. A preliminary mesh sensitivity study was performed, along with a check on the time increments used, in order to ensure that the solution had converged.}.

We proceed to consider 3 representative EP tests A, B and C on a ferritic steel; the steel has a fixed value of trap depth and density. The trap binding energy is fixed at $\Delta\overline{ H}=-13.81$ corresponding to $K={{10}^{6}}$ and the trap density is assumed to be $ \bar{N}={{10}^{-3}}$, implying that $K \bar{N}={{10}^{3}}$. Each test is a three stage EP test and the lattice hydrogen occupancy on the entry face of the EP specimen is taken to be $\theta _{L}^{0}={10}^{-7}, \, {10}^{-4}$ and ${{10}^{-2}}$ for tests A, B and C, respectively. The aim is to show that the tests A to C have significantly different single stage EP responses and likewise have significantly different 3 stage EP responses. Particular emphasis will be placed on the ratio of lag time in stages 1 and 3 for each test A to C.  

\begin{figure}[H]
    \centering
    \begin{subfigure}[b]{0.32\textwidth} 
    \includegraphics[width=\textwidth]{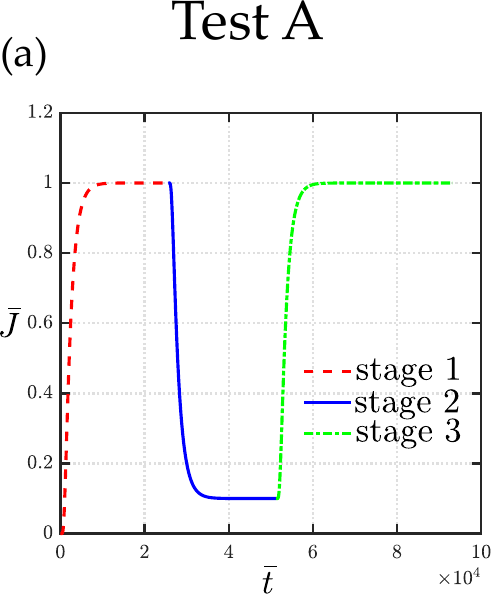} 
    \label{fig:reg1a}
    \end{subfigure} 
    \begin{subfigure}[b]{0.323\textwidth} 
    \includegraphics[width=\textwidth]{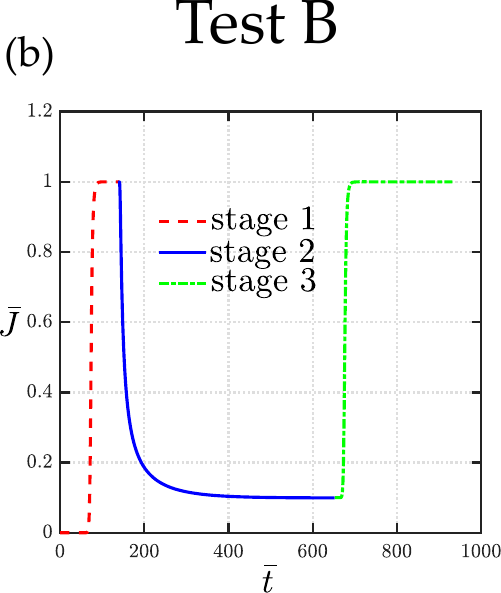}
    \label{fig:reg2a}
    \end{subfigure}
    \begin{subfigure}[b]{0.32\textwidth} 
    \includegraphics[width=\textwidth]{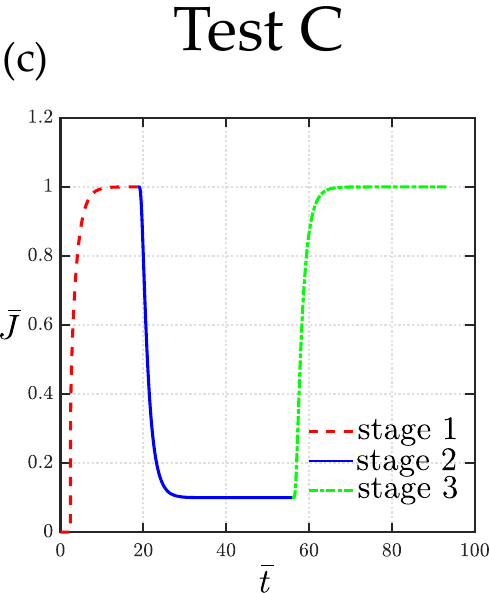}
    \label{fig:reg3a}
    \end{subfigure}    
    
    \begin{subfigure}[b]{0.32\textwidth} 
    \includegraphics[width=\textwidth]{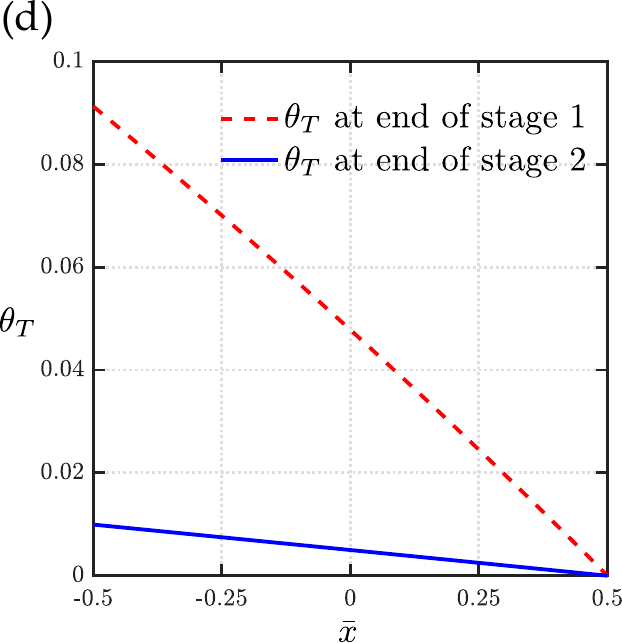} 
    \label{fig:reg1b}
    \end{subfigure}
    \begin{subfigure}[b]{0.32\textwidth} 
    \includegraphics[width=\textwidth]{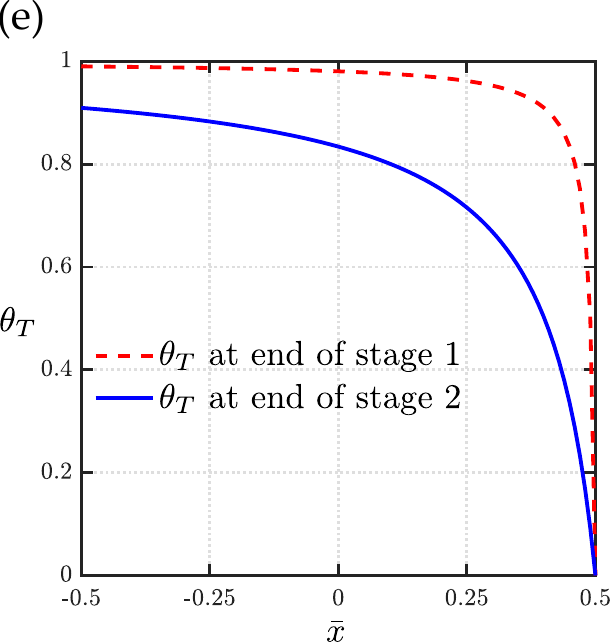}
    \label{fig:reg2b}
    \end{subfigure}
    \begin{subfigure}[b]{0.32\textwidth} 
    \includegraphics[width=\textwidth]{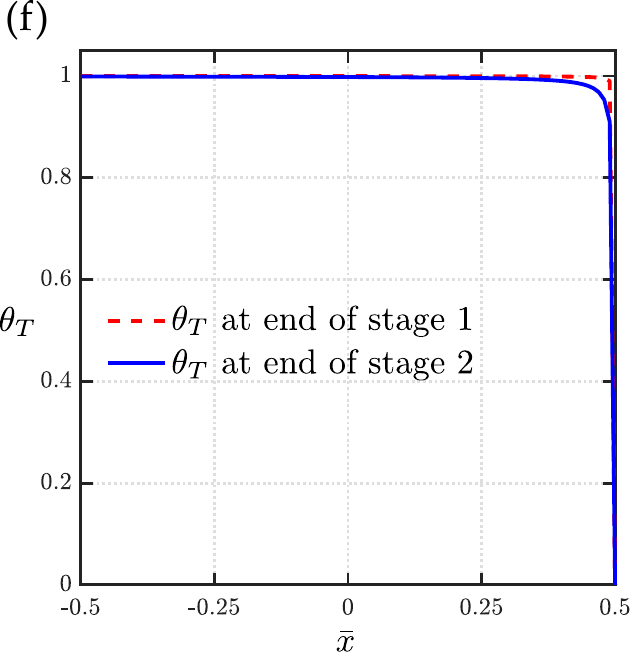}
    \label{fig:reg3b}
    \end{subfigure}      
    \caption{Numerical simulations of three stage EP tests performed in different regimes. The normalised flux \( \bar{J} \) versus time \( \bar{t} \) in three stages for each regime are shown in the top row and the distributions of trap occupancy \( {{\theta }_{T}} \) over the specimen thickness \( 0\le \bar{x}\le 1 \) at the end of stages 1 and 2 are shown in the bottom row. Results of test A are plotted in (a) and (d) where \( \theta_{L}^{0}={{10}^{-7}} \) is used, test B in (b) and (e) where \( \theta_{L}^{0}={{10}^{-4}} \) is used, and test C in (c) and (f) where \( \theta _{L}^{0}={{10}^{-2}} \) is used. In all cases, $K=10^{6}~(\Delta\overline{ H}=-13.81)$, $\bar{N}=10^{-3}$ and $\eta=0.1$.}
    \label{fig5}
\end{figure}

For test A, choose \( \theta _{L}^{0}={{10}^{-7}} \) such that \( K  \theta _{L}^{0}={{10}^{-1}} \), and consequently the response lies within regime I for a single stage EP test.  The small value of  \( K  \theta _{L}^{0}\) implies that the trap occupancy \( \theta _{T} \) is small, and is on the order of \( K  \theta _{L}\). In the 3 stage EP test, take \( \eta=0.1\) for charging of the sample in stage 2. The trap occupancy at the end of stages 1 and 2 remain small, see Fig. \ref{fig5}d, and the efflux in each stage is potted in Fig. \ref{fig5}a. The transient in efflux is comparable for stages 1 and 3, consistent with the fact the trap occupancy is low at the end of stage 2.  

Second, for test B, choose \( \theta _{L}^{0}={{10}^{-4}} \) such that \( K  \theta _{L}^{0}={{10}^{2}} \),  \( {K\bar{N}}/{(K{{\theta _{L}^{0})}^{2}}}=0.1 \) and \( K{\theta_{L}^{0}}/{K\bar{N}}=0.1\). Consequently, the response lies within regime IIIa for a single stage EP test.  In the 3 stage EP test, again with \( \eta=0.1\) for charging of the sample in stage 2, the trap occupancy at the end of stages 1 and 2 is high except near the outlet face of the specimen, see Fig. \ref{fig5}e. Consequently, the duration of the efflux transient in stage 3 is significantly shorter than in stage 1, see  Fig. \ref{fig5}b.

Third, for test C, choose \( \theta _{L}^{0}={{10}^{-2}} \) such that \( K  \theta _{L}^{0}={{10}^{4}} \),  \( {K\bar{N}}/{(K{{\theta _{L}^{0})}^{2}}}={{10}^{-5}} \) and \( K{\theta_{L}^{0}}/{K\bar{N}}=10 \) . Consequently, the response lies within regime IIIb for a single stage EP test.  In this regime, the traps play a negligible role and hydrogen permeation is dictated by Fickian diffusion through the lattice. The  trap occupancy is close to unity, see Fig. \ref{fig5}f, and the effluxes in stage 1 and 3 are similar, see Fig. \ref{fig5}c. 

\begin{figure}[htp]
\centering
\includegraphics[width=0.65\textwidth]{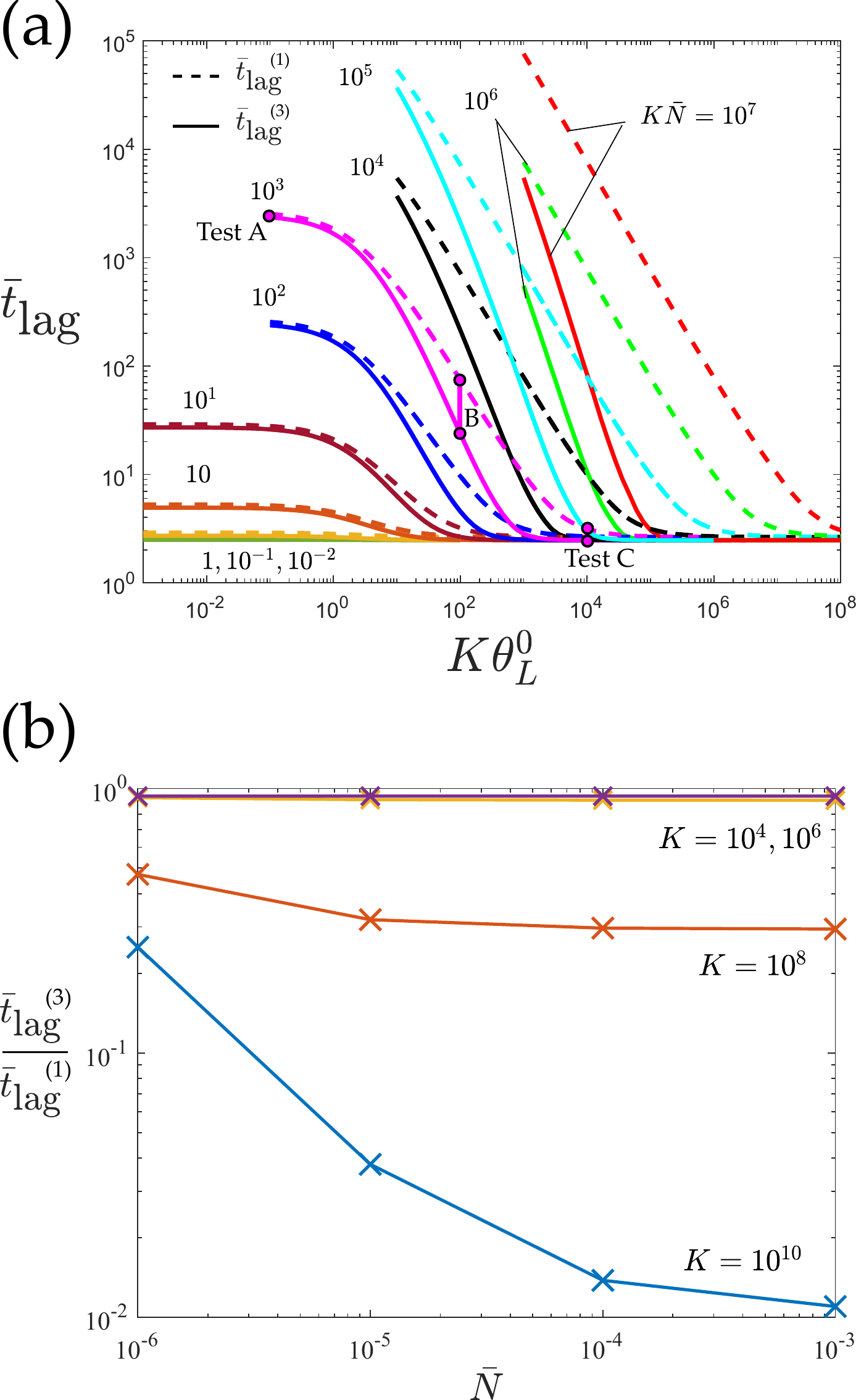}
\caption{(a) Dependence of ${{\bar{t}}_{\text{lag}}}$ upon $K\theta_{L}^{0}$ in a three stage EP test for selected values of ${K\bar{N}}$. Results are shown for both stage 1 (dashed lines) and stage 3 (solid lines). Tests A, B and C have been added as discrete data. (b) The ratio of lag-times in stages 3 and 1 as a function of trap density, for selected values of trap binding energy (expressed in terms of equilibrium constant), for $\theta_L^0=10^{-6}$. These data are a cross-plot from (a) of the figure.}
\label{fig:newfig4}
\end{figure}

The dependence of ${{\bar{t}}_{\text{lag}}}$ upon $K\theta_{L}^{0}$ for stages 1 and 3 of a three stage EP test is plotted is plotted in Fig. \ref{fig:newfig4} for selected values of ${K\bar{N}}$ in a map similar to Fig. \ref{fig4}. The predictions are the full numerical solution to the PDE \eqref{eq9}. The lag time for stage 3 is comparable to that for stage 1 except in an intermediate regime that corresponds to regimes II and IIIa of the asymptotic analysis for stage 1. The 3 tests A, B and C have been added to the figure as discrete data; again, it is clear that test B is the only case where the lag time for diffusion in stage 3 is significantly faster than that in stage 1. The shape of the curves of ${{\bar{t}}_{\text{lag}}}$ versus $K\theta_{L}^{0}$ follows that shown in the sketch of Fig. \ref{fig4}.  However, the full extent of the sigmoidal curves is not shown in all cases in order to emphasise the practical regime of parameter space.

It is very challenging to attempt to come up with an inverse engineering approach to determine the various diffusion and trap parameters from the results of a limited number of 3 stage electro-permeation tests. As a first step in developing such a procedure, the data of Fig. 4a is replotted in the form of the ratio of time lag for stages 3 and 1, for the choice $\theta_L^0=10^{-6}$, see Fig. 4b. The lag time for stage 3 is much less than that in stage 1 when there is a high density of deep traps. Otherwise, the lag times are similar for stages 3 and 1. Thus, a knowledge of the measured ratio of lag times, along with the known density of traps, gives the trap binding energy when a single trap is present.  A similar cross-plot of the data of Fig. 4a can be used for other choices of value of $\theta_L^0$.

\subsection{Case study}\label{sec4a}

Consider a ferritic steel with lattice activation energy $Q=6.7$ kJ mol$^{-1}$, diffusion pre-exponential factor \( {{D}_{0}}=2\times {{10}^{-7}}{{\text{m}}^{\text{2}}}{{\text{s}}^{-1}} \)  and lattice site density \( {{N}_{L}}=8.46\times {{10}^{28}}\text{atoms}\cdot{{\text{m}}^{-3}} \) \cite{novak+etal12}, with \( \alpha=\beta=1 \). Let the test temperature be $T_0=293$ K such that the non-dimensional lattice activation energy is \( \bar{Q}=2.75 \). Assume a physically relevant range of trap binding energy from $-22$ kJ mol$^{-1}$ to $-56$ kJ mol$^{-1}$. The trap density \( {{N}_{T}} \) is taken to be in the range \( ({{10}^{-3}}-{{10}^{-6}}){{N}_{L}} \), hence \( {{10}^{-6}}\le \bar{N}\le {{10}^{-3}} \), as assumed by \cite{oriani70}.

\begin{figure}[H]
    \centering
    \begin{subfigure}[b]{0.45\textwidth} 
    \includegraphics[width=\textwidth]{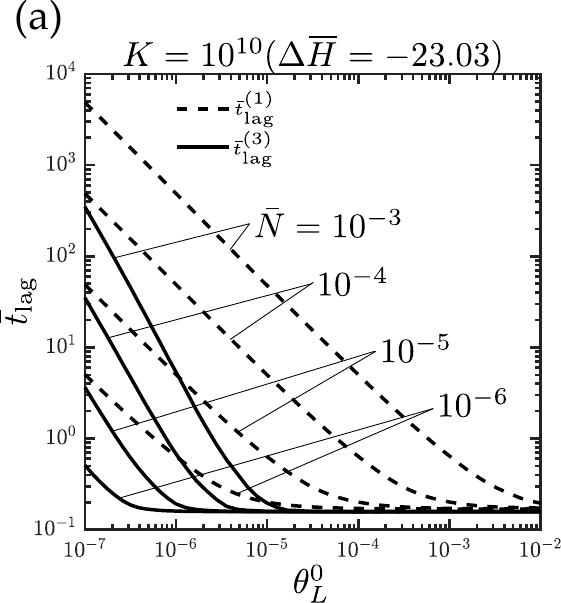} 
    \label{fig:k1e10}
    \end{subfigure} \hspace{10mm}
    \begin{subfigure}[b]{0.45\textwidth} 
    \includegraphics[width=\textwidth]{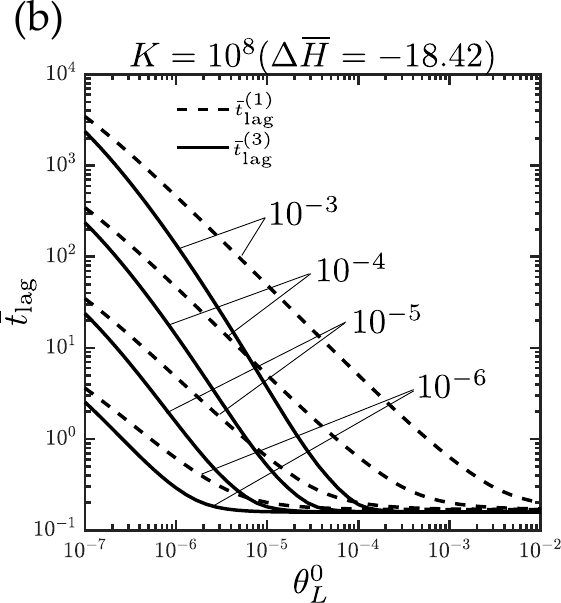}
    \label{fig:k1e8}
    \end{subfigure}
    
    \begin{subfigure}[b]{0.45\textwidth} 
    \includegraphics[width=\textwidth]{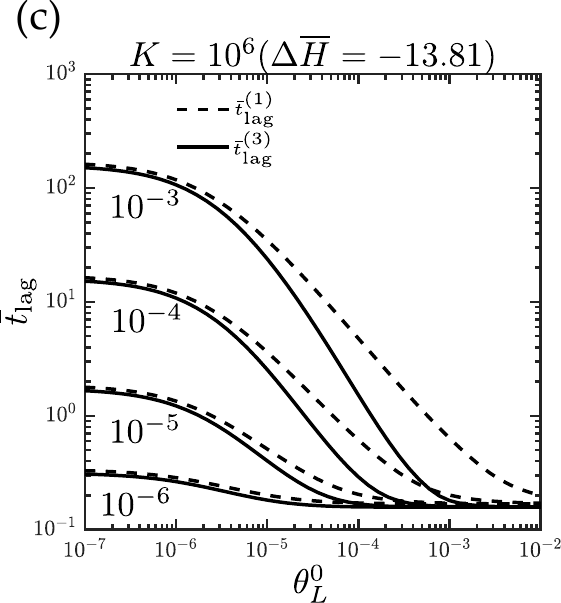} 
    \label{fig:k1e6}
    \end{subfigure} \hspace{10mm}
    \begin{subfigure}[b]{0.45\textwidth} 
    \includegraphics[width=\textwidth]{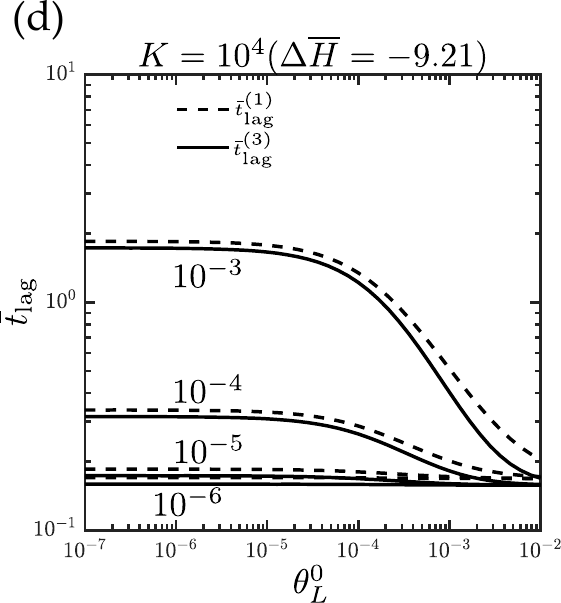}
    \label{fig:k1e4}
    \end{subfigure}    
    \caption{Simulation results of time lag \( \bar{t}_{\text{lag}}^{(1)} \)  and \( \bar{t}_{\text{lag}}^{(3)} \) in stages 1 and 3, respectively, on the $y$-axis are plotted against the initial lattice occupancy fraction \( {{10}^{-7}}\le \theta _{L}^{0}\le {{10}^{-2}} \) on $x$-axis. Results are shown for (a) $K=10^{10}~(\Delta\overline{ H}=-23.03)$, (b) $K=10^{8}~(\Delta\overline{ H}=-18.42)$, (c) $K=10^{6}~(\Delta\overline{ H}=-13.81)$,  and (d) $K=10^{4}~(\Delta\overline{ H}=-9.21)$. In each of the plots, \( \bar{N} \) is varied from \( {{10}^{-6}} \) to \( {{10}^{-3}} \). For all simulations, \( \eta =0.1 \). Depending on the combination of parameters \( \{K,\bar{N},\theta _{L}^{0}\} \), diffusion in stage I may take place in regime I, II or IIIb which can be easily identified by the value of the slope of plots as schematically shown in Fig. \ref{fig4}. Time lag values \( \bar{t}_{\text{lag}}^{(3)} \) in stage 3 resembles the values of lattice diffusion only for deep traps when \( K>{{10}^{8}} \) and \( \theta _{L}^{0}\ge {{10}^{-5}} \) for all quoted values of \( \bar{N} \).}
    \label{fig6}
\end{figure}

In Fig. \ref{fig6}, we explore the permeation transients in the first and third stages of EP over the complete physical range of parameters $\Delta\overline{ H}, \overline{N}$ and \( \theta_{L}^{0} \). As defined in Section \ref{sec2d}, we calculate the time lag values $\bar{t}^{(1)}_{\text{lag}}$ and $\bar{t}^{(3)}_{\text{lag}}$, in stages 1 and 3 respectively, when \( \bar{J}=0.63 \). Limit attention to \( \eta=0.1 \), trap density $\bar{N}\in[10^{-6},\,10^{-3}]$ and trap binding energy $\Delta\overline{ H}\in[-23,\,-9]$ (such that $K\in[10^4,\,10^{10}]$) and perform simulations for $\theta^0_L\in[10^{-7},\,10^{-2}]$. In order to identify the regimes of diffusion, we plot the predictions in terms of \( {{\bar{t}}_{\text{lag}}} \) versus \( \theta_{L}^{0} \), similar to Fig. \ref{fig4}. The results for $K=10^{10}~(\Delta\overline{ H}=-23.03)$ are plotted in Fig. \ref{fig6}a  for the selected values of \( \bar{N} \). The permeation transients in stage 1 reside in regimes II or III; for example, the plot for \( \bar{N}={{10}^{-6}} \) lies in regime II when \( \theta_{L}^{0}<{{10}^{-5}} \) and otherwise in regime IIIb. The time lag values \( \bar{t}_{\text{lag}}^{(3)} \) in stage 3 are orders of magnitude smaller than \( \bar{t}_{\text{lag}}^{(1)} \) at any given \( \bar{N} \). For \( \theta_{L}^{0}\ge {{10}^{-5}} \), permeation transients in stage 3  obey lattice diffusion whereas the corresponding permeation transients in stage 1 are in regime I. Thus, traps of binding energy \( \overline{\Delta H}<-20 \) represent \textit{irreversible} traps which get filled during stage 1 permeation, and lead to lattice diffusion during stage 3.  

In similar fashion, time lag predictions for stages 1 and 3 are obtained for $K=10^{8}~(\Delta\overline{ H}=-18.42)$, $K=10^{6}~(\Delta\overline{ H}=-13.81)$ and $K=10^{4}~(\Delta\overline{ H}=-9.21)$, see  Figs. \ref{fig6}b, \ref{fig6}c and \ref{fig6}d, respectively. It can be observed that, for lower values of trap binding energy \( (\Delta\overline{ H}>-10) \), permeation transients in stages 1 and 3 have similar time lags. This corresponds to the \textit{reversible} trap type, such that hydrogen empty from the traps in stage 2. For the most shallow trap considered, $K=10^{4}~(\Delta\overline{ H}=-9.21)$, hydrogen permeation lies in Regime I, and lattice-driven diffusion is not observed even for the largest $\theta_L^0$ considered.

\begin{figure}[H]
    \centering
    \begin{subfigure}[b]{0.45\textwidth} 
    \includegraphics[width=\textwidth]{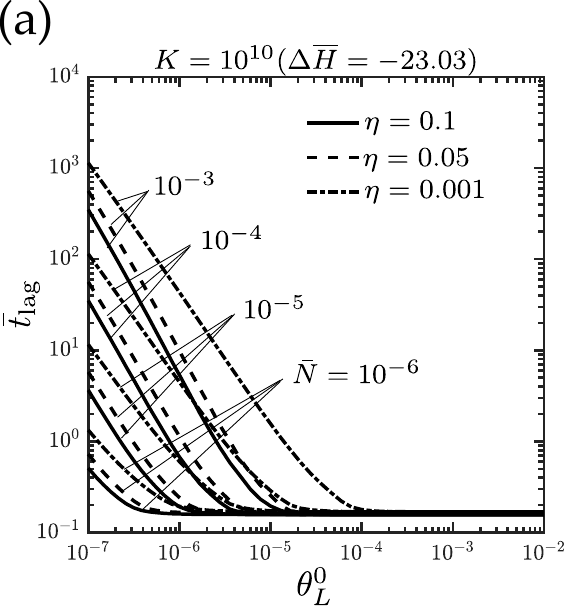} 
    \label{fig:effectEtaK10}
    \end{subfigure} \hspace{10mm}
    \begin{subfigure}[b]{0.45\textwidth} 
    \includegraphics[width=\textwidth]{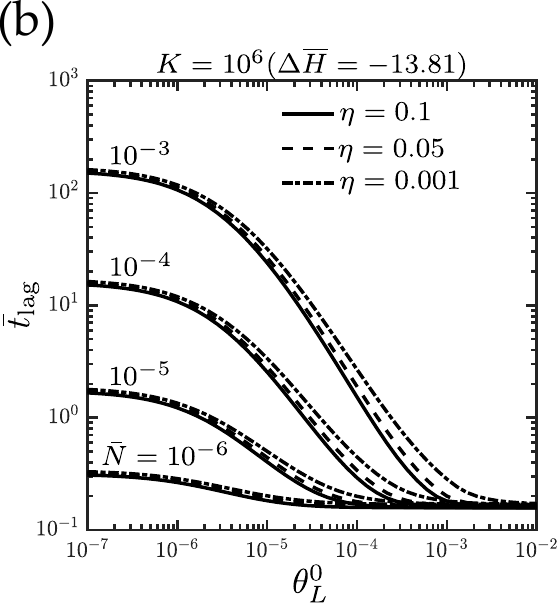}
    \label{fig:effectEtaK6}
    \end{subfigure}
    \caption{Effect of parameter $\eta\in[0.001,\,0.1]$ on time lag values \( \bar{t}_{\text{lag}}^{(3)} \) in stage 3 of EP tests. Results are shown for (a) $K=10^{10}~(\Delta\overline{ H}=-23.03)$  and (b) $K=10^{6}~(\Delta\overline{ H}=-13.81)$ where \( \bar{N} \) is varied from \( {{10}^{-6}} \) to \( {{10}^{-3}} \) in both cases. The effect of \( \eta \) is only significant for deep traps $\Delta\overline{ H}\le-23.03~(\Delta H\le-56\text{kJ/mol})$.}
    \label{fig7}
\end{figure}

Finally, we investigate the role of the parameter $\eta$ which characterises the magnitude of the background current in stage 2. The simulations of Figs. \ref{fig6}a and \ref{fig6}c for \( \eta =0.1 \) are repeated for \( \eta =0.05 \) and $0.001$  in Fig. \ref{fig7}. The results show that it takes longer to achieve  steady state in stage 2 of the EP test with decreasing \( \eta \): more hydrogen diffuses out of traps in stage 2 when \( \eta \) is reduced.  Consequently, the time lag for permeation of flux in stage 3 increases with decreasing \( \eta \). In broad terms, the sensitivity of the stage 3 transient to the value of \( \eta \) is greatest for deep traps (high binding energy).

\section{Conclusions}
We have presented a numerical analysis of the three stage electro-permeation (EP) experiment. Stage 1 is characterised by a rising permeation flux until steady state is achieved for the given charging condition. In stage 2, the input charging condition is reduced to a fraction of the stage 1 charging condition so that hydrogen is released from both the lattice and trap sites of low binding energy. Stage 2 is immediately followed by stage 3 where the charging conditions of stage 1 are reimposed. The difference between the rising permeation fluxes in stages 1 and 3 stems from the level of trap occupancy that persists at the end of stage 2. The analysis is framed in terms of regimes of behaviour, as characterised by the lag times. A universal map to give insight into the relative response of stages 1 and 3 has been generated.  The regimes of the map are in good agreement with analytical expressions for diffusion rate in the first stage of a 3  stage EP test.

\begin{itemize}
    \item  In regime I, most of  hydrogen that has been trapped in stage 1 exhausts from the outlet face at the end of stage 2. In  contrast, in regimes II and III only a small fraction of the trapped hydrogen has diffused out at the end of stage 2.
    
    \item Deep traps ($\Delta\overline{ H}<-20$) are filled during stage 1 and do not interfere with lattice diffusion during stage 3. In contrast, traps of $\Delta\overline{ H}>-20$ release hydrogen into the lattice during stage 2.
    
    \item Stage 3 time lags can be orders of magnitude smaller than those of stage 1 in the presence of deep traps, thereby facilitating the quantification of the lattice diffusivity from the stage 3 response. However, for traps of low and medium binding energies, the differences between the responses in the first and third stages of a three stage EP test are small.

\end{itemize}

\section{Acknowledgments}
\label{Acknowledge of funding}

 E. Mart\'{\i}nez-Pa\~neda acknowledges financial support from EPSRC [grant EP/V009680/1] and UKRI's Future Leaders Fellowship programme [grant MR/V024124/1].




\bibliographystyle{elsarticle-num}
\bibliography{references}

\end{document}